\documentclass[12pt,a4paper]{article}
\usepackage{amsmath}
\usepackage{latexsym}
\makeatletter
\def\rddots{\mathinner{\mkern1mu\raise\p@%
    \vbox{\kern7\p@\hbox{.}}\mkern2mu%
    \raise4\p@\hbox{.}\mkern2mu\raise7\p@\hbox{.}\mkern1mu}}
\makeatother
\newcommand{\fukuso}{{\mathbf C}}

\begin{document}

\title{\sl A Geometric Parametrization of the Cabibbo--
Kobayashi--Maskawa Matrix and the Jarlskog Invariant}
\author{
  Kazuyuki FUJII
  \thanks{E-mail address : fujii@yokohama-cu.ac.jp }\\
  Department of Mathematical Sciences\\
  Yokohama City University\\
  Yokohama, 236--0027\\
  Japan
  }
\date{}
\maketitle
\begin{abstract}
  In this paper we give a geometric parametrization to the Cabibbo--
  Kobayashi--Maskawa (CKM) mixing matrix and the Jarlskog invariant, 
  which is based on two flag manifolds $SU(3)/U(1)^{2}$.

  To treat a fourth generation of quarks on CP violation we  
  generalize the parametrization to one based on two flag manifolds 
  $SU(4)/U(1)^{3}$.
\end{abstract}
%


%
%
%
%
\newpage

\section{Introduction}
CP violation plays a central role in the standard model. In this paper 
we revisit a problem of generation of quarks on CP violation 
(\cite{Ca}, \cite{KM}, \cite{Ja1}) from the mathematical (geometric) point 
of view. As a quick introduction to the problem see for example \cite{CG}. 

We start with the paper \cite{Ja1} and assume for simplicity that the mass 
matrices $M$ and $M^{\prime}$ are hermite and non-negative. 
Then $M$ and $M^{\prime}$ can be diagonalized like
\begin{equation}
\label{eq:mass}
M=
U
\left(
  \begin{array}{ccc}
    m_{u} &       &        \\
          & m_{c} &        \\
          &       & m_{t} 
  \end{array}
\right)
U^{\dagger},
\qquad
M^{\prime}=
U^{\prime}
\left(
  \begin{array}{ccc}
    m_{d} &       &        \\
          & m_{s} &        \\
          &       & m_{b} 
  \end{array}
\right)
{U^{\prime}}^{\dagger}
\end{equation}
where $m_{j}$ is the mass of the quark $j$. 
From these we define
\begin{equation}
\label{CKM}
V
=U^{\dagger}U^{\prime}
=
\left(
  \begin{array}{ccc}
    V_{ud} & V_{us} & V_{ub} \\
    V_{cd} & V_{cs} & V_{cb} \\
    V_{td} & V_{ts} & V_{tb}
  \end{array}
\right)
\equiv 
\left(
  \begin{array}{ccc}
    V_{11} & V_{12} & V_{13} \\
    V_{21} & V_{22} & V_{23} \\
    V_{31} & V_{32} & V_{33}
  \end{array}
\right),
\end{equation}
which is the famous Cabibbo--Kobayashi--Maskawa (CKM) mixing 
matrix.

The Jarlskog invariant which measures CP violation is given by 
\begin{equation}
\label{eq:invariant}
J=-i\ \det[M,M^{\prime}]/2TB
\end{equation}
where
\[
T=(m_{t}-m_{u})(m_{t}-m_{c})(m_{c}-m_{u}),\quad 
B=(m_{b}-m_{d})(m_{b}-m_{s})(m_{s}-m_{d}).
\]
In terms of entries of $V$ in (\ref{CKM}) $J$ is expressed as
\begin{equation}
\label{eq:V-expression}
J=\pm \mbox{Im}\ (V_{ij}V_{kl}\bar{V}_{il}\bar{V}_{kj})
\end{equation}
where Im denotes the imaginary part of complex number and 
$\bar{z}$ the complex conjugate of $z$. See the appendix. 
Usually $\mbox{Im}\ (V_{11}V_{22}\bar{V}_{12}\bar{V}_{21})$ is used, 
while $\mbox{Im}\ (V_{11}V_{33}\bar{V}_{13}\bar{V}_{31})$ is used 
in the following. 

From (\ref{eq:mass}) the forms are invariant under
\[ 
U\ \rightarrow\ U\ \mbox{diag}(e^{i\theta_{1}},e^{i\theta_{2}},
e^{i\theta_{3}}),
\quad
U^{\prime}\ \rightarrow\ U^{\prime}\ 
\mbox{diag}(e^{i{\theta_{1}}^{\prime}},e^{i{\theta_{2}}^{\prime}},
e^{i{\theta_{3}}^{\prime}}),
\]
so $U$ and $U^{\prime}$ are considered as elements in a flag 
manifold $U(3)/U(1)^{3}\cong SU(3)/U(1)^{2}$.

Since the dimension of $SU(3)/U(1)^{2}$ is six, $U$ is usually 
parametrized as
\begin{equation}
\label{eq:old-parametrization}
U=
\left(
  \begin{array}{ccc}
    e^{i(\alpha +\beta)} &                      &              \\
                         & e^{i(\alpha -\beta)} &              \\
                         &                      & e^{-2i\alpha}
  \end{array}
\right)
\left(
  \begin{array}{ccc}
    c_{12}c_{13} & s_{12}c_{13} & s_{13}e^{-i\delta} \\
   -s_{12}c_{23}-c_{12}s_{23}s_{13}e^{i\delta} & 
    c_{12}c_{23}-s_{12}s_{23}s_{13}e^{i\delta} & s_{23}c_{13} \\
    s_{12}s_{23}-c_{12}c_{23}s_{13}e^{i\delta} & 
   -c_{12}s_{23}-s_{12}c_{23}s_{13}e^{i\delta} & c_{23}c_{13}
  \end{array}
\right)
\end{equation}
where $c_{ij}=\cos\theta_{ij}$, $s_{ij}=\sin\theta_{ij}$ and 
$\{\theta_{12},\theta_{13},\theta_{23}\}$ are three rotating angles 
and $e^{i\delta}$ is a phase, see for example \cite{GGPT} and \cite{PDG}. 

However, we don't use this parametrization. Since $SU(3)/U(1)^{2}$ is 
a K{\" a}hler manifold, there is some deep geometric structure. 
As a general introduction to Geometry or Topology see \cite{Na} and 
\cite{Fu1} (\cite{Na} is particularly recommended).

Our aim is to give a geometric parametrization to $V$ by use of a 
local coordinate of two flag manifolds $SU(3)/U(1)^{2}$ 
corresponding to $U$ and $U^{\prime}$ (see the following diagram).

\vspace{5mm}
\begin{center}
\input{CKM.fig}
\end{center}

\vspace{5mm}
We note that the argument above is based on the third generation of 
quarks, but it is easy to generalize to any generation except for 
calculation. In fact, we treat a fourth generation and consider 
two flag manifolds $SU(4)/U(1)^{3}$ in the following.

\section{Geometric Parametrization}
First of all we review how to parametrize the flag manifold 
$SU(3)/U(1)^{2}\cong U(3)/U(1)^{3}$. See \cite{Pi} as 
a good introduction and also \cite{DJ} and \cite{FO}.

The flag manifold of the second type (in our terminology) 
is the sequence of complex vector spaces defined by
\begin{equation}
F_{1,1,1}(\fukuso)=\{{\cal V}\subset {\cal W}\subset \fukuso^{3}\ 
|\ \mbox{dim}_{\fukuso}{\cal V}=1,\ \mbox{dim}_{\fukuso}{\cal W}=2\}.
\end{equation}
Then it is well--known that
\begin{equation}
F_{1,1,1}(\fukuso)\cong U(3)/U(1)^{3}
\end{equation}
and moreover
\begin{equation}
U(3)/U(1)^{3}\cong GL(3;\fukuso)/B_{+}
\end{equation}
where $B_{+}$ is the (upper) Borel subgroup given by
\[
B_{+}=
\left\{
\left(
\begin{array}{ccc}
\alpha & *     & *      \\
0      & \beta & *      \\
0      & 0     & \gamma 
\end{array}
\right)
\in GL(3;\fukuso)\ |\ \alpha,\ \beta,\ \gamma \in 
GL(1;\fukuso)\equiv \fukuso^{\times}
\right\}.
\]

In order to obtain the element of $U(3)/U(1)^{3}$ 
from an element in $GL(3;\fukuso)/B_{+}$ it is convenient to use 
the orthonormalization (method) by Gram--Schmidt. 
For the matrix
\begin{equation}
F\equiv
\left(
\begin{array}{ccc}
1 & 0 & 0 \\
x & 1 & 0 \\
y & z & 1
\end{array}
\right)
\ \in\ GL(3;\fukuso)/B_{+}
\end{equation}
we set
\[
V_{1}
=
\left(
\begin{array}{c}
1 \\
x \\
y
\end{array}
\right),\quad
V_{2}
=
\left(
\begin{array}{c}
0 \\
1 \\
z
\end{array}
\right),\quad
V_{3}
=
\left(
\begin{array}{c}
0 \\
0 \\
1
\end{array}
\right).
\]

\vspace{5mm}
For $\{V_{1},V_{2},V_{3}\}$ the Gramm--Schmidt orthogonalization 
is as follows :
\begin{eqnarray*}
\tilde{V}_{1}&=&V_{1},\quad 
\hat{V}_{1}=\tilde{V}_{1}(\tilde{V}_{1}^{\dagger}\tilde{V}_{1})^{-1/2}
\ \Longrightarrow\ 
P_{1}=\hat{V}_{1}{\hat{V}_{1}}^{\dagger}\ :\ \mbox{projection}  \\
\tilde{V}_{2}&=&(E-P_{1})V_{2},\quad 
\hat{V}_{2}=\tilde{V}_{2}(\tilde{V}_{2}^{\dagger}\tilde{V}_{2})^{-1/2}
\ \Longrightarrow\ 
P_{2}=\hat{V}_{2}{\hat{V}_{2}}^{\dagger}\ :\ \mbox{projection}  \\
\tilde{V}_{3}&=&
(E-P_{1}-P_{2})V_{3}=(E-P_{1})(E-P_{2})V_{3}
,\quad 
\hat{V}_{3}=\tilde{V}_{3}(\tilde{V}_{3}^{\dagger}\tilde{V}_{3})^{-1/2}
\end{eqnarray*}
where $E$ is the unit matrix in $M(3;\fukuso)$. Explicitly,
\[
\hat{V}_{1}=
\left(
\begin{array}{c}
1 \\
x \\
y
\end{array}
\right) {\Delta_{1}}^{-1/2},\quad
\hat{V}_{2}=
\left(
\begin{array}{c}
-(\bar{x}+\bar{y}z) \\
1-(xz-y)\bar{y}     \\
z+\bar{x}(xz-y)
\end{array}
\right) 
{(\Delta_{1}\Delta_{2})}^{-1/2},\quad
\hat{V}_{3}=
\left(
\begin{array}{c}
\bar{x}\bar{z}-\bar{y} \\
-\bar{z}               \\
1           
\end{array}
\right)
{\Delta_{2}}^{-1/2}
\]
where
\[
\Delta_{1}=1+|x|^{2}+|y|^{2},\quad
\Delta_{2}=1+|z|^{2}+|xz-y|^{2}.
\]
Note that $K=\log(\Delta_{1}\Delta_{2})$ is the K{\"a}hler potential 
and $\omega=i\partial\bar{\partial}K$ is the K{\"a}hler two--form 
and the symplectic volume of the manifold is given by
\begin{equation}
\Omega\equiv \frac{\omega\wedge\omega\wedge\omega}{3!}=
\frac{2}{(\Delta_{1}\Delta_{2})^{2}}
\prod_{j=1}^{3}idz_{j}\wedge d\bar{z}_{j}
\end{equation}
where we have set $z_{1}=x,\ z_{2}=y,\ z_{3}=z$ for simplicity, 
see for example \cite{Pi}.

Therefore we obtain the unitary matrix
\begin{eqnarray}
\label{eq:new-parametrization}
U=(\hat{V}_{1},\hat{V}_{2},\hat{V}_{3})
&=&
\left(
\begin{array}{ccc}
\frac{1}{\sqrt{\Delta_{1}}} & 
\frac{-(\bar{x}+\bar{y}z)}{\sqrt{\Delta_{1}\Delta_{2}}} &
\frac{\bar{x}\bar{z}-\bar{y}}{\sqrt{\Delta_{2}}}            \\
\frac{x}{\sqrt{\Delta_{1}}} & 
\frac{1-(xz-y)\bar{y}}{\sqrt{\Delta_{1}\Delta_{2}}} &
\frac{-\bar{z}}{\sqrt{\Delta_{2}}}                          \\
\frac{y}{\sqrt{\Delta_{1}}} &
\frac{z+\bar{x}(xz-y)}{\sqrt{\Delta_{1}\Delta_{2}}} &
\frac{1}{\sqrt{\Delta_{2}}}
\end{array}
\right)  \nonumber \\
&=&
\left(
\begin{array}{ccc}
1 & -(\bar{x}+\bar{y}z) & \bar{x}\bar{z}-\bar{y} \\
x & 1-(xz-y)\bar{y}     & -\bar{z}               \\
y & z+\bar{x}(xz-y)     & 1
\end{array}
\right)
\left(
\begin{array}{ccc}
\frac{1}{\sqrt{\Delta_{1}}} &  &                     \\
          & \frac{1}{\sqrt{\Delta_{1}\Delta_{2}}} &  \\
          &         & \frac{1}{\sqrt{\Delta_{2}}}
\end{array}
\right).
\end{eqnarray}
This is our geometric parametrization for $U$.

\vspace{3mm}
A comment is in order.\ Our parametrization is not compatible with 
(\ref{eq:old-parametrization}) because of the phases in the first. 
However, if we neglect them the correspondence is given by
\[
x\ \longleftrightarrow\ 
-\left(t_{12}\frac{c_{23}}{c_{13}}+s_{23}t_{13}e^{i\delta}\right),\quad
y\ \longleftrightarrow\ 
\left(t_{12}\frac{s_{23}}{c_{13}}-c_{23}t_{13}e^{i\delta}\right),\quad
z\ \longleftrightarrow\
-t_{23}
\]
where $c_{ij}=\cos\theta_{ij},\ s_{ij}=\sin\theta_{ij},\ 
t_{ij}=\tan\theta_{ij}$ \ for simplicity.

\vspace{5mm}
Similarly, we parametrize $U^{\prime}$ in terms of 
$(u,v,w)$ (in place of $(x,y,z)$ in $U$) as
\begin{equation}
U^{\prime}=
\left(
\begin{array}{ccc}
1 & -(\bar{u}+\bar{v}w) & \bar{u}\bar{w}-\bar{v} \\
u & 1-(uw-y)\bar{v}     & -\bar{w}               \\
v & w+\bar{u}(uw-v)     & 1
\end{array}
\right)
\left(
\begin{array}{ccc}
\frac{1}{\sqrt{\Delta_{1}^{\prime}}} &  &                     \\
      & \frac{1}{\sqrt{\Delta_{1}^{\prime}\Delta_{2}^{\prime}}} & \\
      &         & \frac{1}{\sqrt{\Delta_{2}^{\prime}}}
\end{array}
\right)
\end{equation}
where 
\[
\Delta_{1}^{\prime}=1+|u|^{2}+|v|^{2},\quad
\Delta_{2}^{\prime}=1+|w|^{2}+|uw-v|^{2}.
\]

Therefore the CKM matrix $V=U^{\dagger}U^{\prime}$ in (\ref{CKM}) 
is parametrized as
\begin{eqnarray}
V&=&
\left(
\begin{array}{ccc}
\frac{1}{\sqrt{\Delta_{1}}} &  &                     \\
          & \frac{1}{\sqrt{\Delta_{1}\Delta_{2}}} &  \\
          &         & \frac{1}{\sqrt{\Delta_{2}}}
\end{array}
\right)
\left(
\begin{array}{ccc}
1 & -(\bar{x}+\bar{y}z) & \bar{x}\bar{z}-\bar{y} \\
x & 1-(xz-y)\bar{y}     & -\bar{z}               \\
y & z+\bar{x}(xz-y)     & 1
\end{array}
\right)^{\dagger}
\left(
\begin{array}{ccc}
1 & -(\bar{u}+\bar{v}w) & \bar{u}\bar{w}-\bar{v} \\
u & 1-(uw-y)\bar{v}     & -\bar{w}               \\
v & w+\bar{u}(uw-v)     & 1
\end{array}
\right) \nonumber \\
&\times&
\left(
\begin{array}{ccc}
\frac{1}{\sqrt{\Delta_{1}^{\prime}}} &  &                     \\
      & \frac{1}{\sqrt{\Delta_{1}^{\prime}\Delta_{2}^{\prime}}} & \\
      &         & \frac{1}{\sqrt{\Delta_{2}^{\prime}}}
\end{array}
\right) \nonumber \\
&=&
\left(
\begin{array}{ccc}
\frac{1}{\sqrt{\Delta_{1}}} &  &                     \\
          & \frac{1}{\sqrt{\Delta_{1}\Delta_{2}}} &  \\
          &         & \frac{1}{\sqrt{\Delta_{2}}}
\end{array}
\right)
\left(
  \begin{array}{ccc}
    f_{11} & f_{12} & f_{13} \\
    f_{21} & f_{22} & f_{23} \\
    f_{31} & f_{32} & f_{33}
  \end{array}
\right)
\left(
\begin{array}{ccc}
\frac{1}{\sqrt{\Delta_{1}^{\prime}}} &  &                     \\
      & \frac{1}{\sqrt{\Delta_{1}^{\prime}\Delta_{2}^{\prime}}} & \\
      &         & \frac{1}{\sqrt{\Delta_{2}^{\prime}}}
\end{array}
\right)
\end{eqnarray}
where
\begin{eqnarray*}
f_{11}&=&1+\bar{x}u+\bar{y}v,\\ 
f_{12}&=&-(\bar{u}+\bar{v}w)+\bar{x}\{1-(uw-v)\bar{v}\}
       +\bar{y}\{w+\bar{u}(uw-v)\},\\      
f_{13}&=&(\bar{u}\bar{w}-\bar{v})-\bar{x}\bar{w}+\bar{y}, \\
f_{21}&=&-(x+y\bar{z})+\{1-(\bar{x}\bar{z}-\bar{y})y\}u
         +\{\bar{z}+x(\bar{x}\bar{z}-\bar{y})\}v,\\
f_{22}&=&(x+y\bar{z})(\bar{u}+\bar{v}w)+
         \{1-(\bar{x}\bar{z}-\bar{y})y\}\{1-(uw-v)\bar{v}\}+
         \{\bar{z}+x(\bar{x}\bar{z}-\bar{y})\}\{w+\bar{u}(uw-v)\},\\
f_{23}&=&-(x+y\bar{z})(\bar{u}\bar{w}-\bar{v})-
         \{1-(\bar{x}\bar{z}-\bar{y})y\}\bar{w}+
         \bar{z}+x(\bar{x}\bar{z}-\bar{y}),\\
f_{31}&=&(xz-y)-zu+v,\\
f_{32}&=&-(xz-y)(\bar{u}+\bar{v}w)-z\{1-(uw-v)\bar{v}\}+
         w+\bar{u}(uw-v),\\
f_{33}&=&(xz-y)(\bar{u}\bar{w}-\bar{v})+z\bar{w}+1.
\end{eqnarray*}

\par \noindent
This is just our geometric parametrization to the CKM matrix. 
We believe that our parametrization is clear--cut.

From this it is easy to see that the Jarlskog invariant in 
(\ref{eq:V-expression}) becomes
\begin{eqnarray}
J
&=&\mbox{Im}\ (V_{11}V_{33}\bar{V}_{13}\bar{V}_{31}) \nonumber \\
&=&
\frac{
\mbox{Im}\left\{
(1+\bar{x}u+\bar{y}v)
(1+z\bar{w}+(xz-y)(\bar{u}\bar{w}-\bar{v}))
(\bar{x}\bar{z}-\bar{y}-\bar{z}\bar{u}+\bar{v})
(uw-v-xw+y)
\right\}
}
{\Delta_{1}\Delta_{2}\Delta_{1}^{\prime}\Delta_{2}^{\prime}}. \nonumber \\
&{}&
\end{eqnarray}
We can of course expand the numerator of the equation. However, such a form 
is not so beautiful, so we omit it.

\section{Generalization to a fourth generation}
In the preceding section we studied some problems of the third generation 
of quarks. However, from the mathematical point of view there is no 
reason to stay at the point (situation). 
Therefore we try to generalize some results based on the third generation 
to ones based on a fourth generation of quarks. 

The method is almost same. Namely, we have only to consider 
a flag manifold $SU(4)/U(1)^{3}\cong U(4)/U(1)^{4}$ in place of 
$SU(3)/U(1)^{2}$ in the preceding section (see the following diagram).

\vspace{5mm}
\begin{center}
\input{CKM-2.fig}
\end{center}

\vspace{5mm}
In order to obtain the element of $SU(4)/U(1)^{3}$ from an element 
in $GL(4;\fukuso)/B_{+}\ (\cong U(4)/U(1)^{4})$ we consider the matrix
\begin{equation}
F\equiv
\left(
\begin{array}{cccc}
1     & 0     & 0     & 0  \\
x_{1} & 1     & 0     & 0  \\
x_{2} & y_{1} & 1     & 0  \\
x_{3} & y_{2} & z_{1} & 1
\end{array}
\right)
\ \in\ GL(4;\fukuso)/B_{+}
\end{equation}
and set
\[
V_{1}
=
\left(
\begin{array}{c}
1     \\
x_{1} \\
x_{2} \\
x_{3} 
\end{array}
\right),\quad
V_{2}
=
\left(
\begin{array}{c}
0     \\
1     \\
y_{1} \\
y_{2}
\end{array}
\right),\quad
V_{3}
=
\left(
\begin{array}{c}
0    \\
0    \\
1    \\
z_{1}
\end{array}
\right),\quad
V_{4}
=
\left(
\begin{array}{c}
0 \\
0 \\
0 \\
1
\end{array}
\right).
\]

\vspace{5mm}
For $\{V_{1},V_{2},V_{3},V_{4}\}$ the Gramm--Schmidt orthogonalization 
is as follows :
\begin{eqnarray*}
&&\tilde{V}_{1}=V_{1},\
\hat{V}_{1}=\tilde{V}_{1}(\tilde{V}_{1}^{\dagger}\tilde{V}_{1})^{-1/2}
\ \Longrightarrow\ 
P_{1}=\hat{V}_{1}{\hat{V}_{1}}^{\dagger}\ :\ \mbox{projection}  \\
&&\tilde{V}_{2}=(E-P_{1})V_{2},\
\hat{V}_{2}=\tilde{V}_{2}(\tilde{V}_{2}^{\dagger}\tilde{V}_{2})^{-1/2}
\ \Longrightarrow\ 
P_{2}=\hat{V}_{2}{\hat{V}_{2}}^{\dagger}\ :\ \mbox{projection}  \\
&&\tilde{V}_{3}=
(E-P_{1}-P_{2})V_{3}=(E-P_{1})(E-P_{2})V_{3}
,\
\hat{V}_{3}=\tilde{V}_{3}(\tilde{V}_{3}^{\dagger}\tilde{V}_{3})^{-1/2}
\ \Longrightarrow\ 
P_{3}=\hat{V}_{3}{\hat{V}_{3}}^{\dagger}\ :\ \mbox{projection}  \\
&&\tilde{V}_{4}=
(E-P_{1}-P_{2}-P_{3})V_{4}=(E-P_{1})(E-P_{2})(E-P_{3})V_{4}
,\
\hat{V}_{4}=\tilde{V}_{4}(\tilde{V}_{4}^{\dagger}\tilde{V}_{4})^{-1/2}
\end{eqnarray*}
where $E$ is the unit matrix in $M(4;\fukuso)$. 

We list the result (whose proof is not easy\footnote{
to calculate the norms $|\tilde{V}_{j}|^{2}\ (j=2,3,4)$ is hard})
\begin{eqnarray*}
\hat{V}_{1}&=&
\left(
\begin{array}{c}
1     \\
x_{1} \\
x_{2} \\
x_{3} 
\end{array}
\right)\frac{1}{\sqrt{\Delta_{1}}},\\
\hat{V}_{2}&=&
\left(
\begin{array}{c}
-T     \\
\Delta_{1}-x_{1}T \\
y_{1}\Delta_{1}-x_{2}T \\
y_{2}\Delta_{1}-x_{3}T 
\end{array}
\right)\frac{1}{\sqrt{\Delta_{1}\Delta_{2}}}
=
\left(
\begin{array}{c}
-(\bar{x}_{1}+\bar{x}_{2}y_{1}+\bar{x}_{3}y_{2}) \\
1+\bar{x}_{2}(x_{2}-x_{1}y_{1})+\bar{x}_{3}(x_{3}-x_{1}y_{2}) \\
y_{1}-\bar{x}_{1}(x_{2}-x_{1}y_{1})-\bar{x}_{3}(x_{2}y_{2}-x_{3}y_{1}) \\
y_{2}-\bar{x}_{1}(x_{3}-x_{1}y_{2})+\bar{x}_{2}(x_{2}y_{2}-x_{3}y_{1})
\end{array}
\right)\frac{1}{\sqrt{\Delta_{1}\Delta_{2}}}
\end{eqnarray*}
where
\[
T=\bar{x}_{1}+\bar{x}_{2}y_{1}+\bar{x}_{3}y_{2},
\]
and
\[
\hat{V}_{3}=
\left(
\begin{array}{c}
a_{1} \\
a_{2} \\
a_{3} \\
a_{4} 
\end{array}
\right)\frac{1}{\Delta_{1}\sqrt{\Delta_{2}\Delta_{3}}}
\]
where
\begin{eqnarray*}
a_{1}&=&-(\bar{x}_{2}+z_{1}\bar{x}_{3})\Delta_{2}+
\{(\bar{y}_{1}\Delta_{1}-\bar{x}_{2}\bar{T})+
  z_{1}(\bar{y}_{2}\Delta_{1}-\bar{x}_{3}\bar{T})\}T, \\
a_{2}&=&-(\bar{x}_{2}+z_{1}\bar{x}_{3})x_{1}\Delta_{2}-
\{(\bar{y}_{1}\Delta_{1}-\bar{x}_{2}\bar{T})+
  z_{1}(\bar{y}_{2}\Delta_{1}-\bar{x}_{3}\bar{T})\}(\Delta_{1}-x_{1}T), \\
a_{3}&=&\Delta_{1}\Delta_{2}-(\bar{x}_{2}+z_{1}\bar{x}_{3})x_{2}\Delta_{2}-
\{(\bar{y}_{1}\Delta_{1}-\bar{x}_{2}\bar{T})+
  z_{1}(\bar{y}_{2}\Delta_{1}-\bar{x}_{3}\bar{T})\}(y_{1}\Delta_{1}-x_{2}T), \\
a_{4}&=&z_{1}\Delta_{1}\Delta_{2}-(\bar{x}_{2}+z_{1}\bar{x}_{3})x_{3}\Delta_{2}-
\{(\bar{y}_{1}\Delta_{1}-\bar{x}_{2}\bar{T})+
  z_{1}(\bar{y}_{2}\Delta_{1}-\bar{x}_{3}\bar{T})\}(y_{2}\Delta_{1}-x_{3}T),
\end{eqnarray*}
and
\[
\hat{V}_{4}=
\left(
\begin{array}{c}
-\bar{x}_{3}+\bar{x}_{1}\bar{y}_{2}+\bar{x}_{2}\bar{z}_{1}
-\bar{x}_{1}\bar{y}_{1}\bar{z}_{1} \\
-\bar{y}_{2}+\bar{y}_{1}\bar{z}_{1} \\
-\bar{z}_{1} \\
1
\end{array}
\right)\frac{1}{\sqrt{\Delta_{3}}}
\equiv
\left(
\begin{array}{c}
b_{1} \\
b_{2} \\
b_{3} \\
b_{4} 
\end{array}
\right)\frac{1}{\sqrt{\Delta_{3}}}.
\]
Here we have used the notations
\begin{eqnarray*}
\Delta_{1}&=&1+|x_{1}|^{2}+|x_{2}|^{2}+|x_{3}|^{2}, \\
\Delta_{2}&=&1+|y_{1}|^{2}+|y_{2}|^{2}+|x_{2}-x_{1}y_{1}|^{2}+
|x_{3}-x_{1}y_{2}|^{2}+|x_{2}y_{2}-x_{3}y_{1}|^{2}, \\
\Delta_{3}&=&1+|z_{1}|^{2}+|y_{2}-y_{1}z_{1}|^{2}+
|x_{1}(y_{2}-y_{1}z_{1})-(x_{3}-x_{2}z_{1})|^{2}.
\end{eqnarray*}
Therefore we have the unitary matrix parametrized by $(x_{1},x_{2},
x_{3},y_{1},y_{2},z_{1})$
\begin{eqnarray}
\label{eq:}
U
&=&(\hat{V}_{1},\hat{V}_{2},\hat{V}_{3},\hat{V}_{4}) \nonumber \\
&=&
\left(
\begin{array}{cccc}
1     & -T                     & a_{1} & b_{1} \\
x_{1} & \Delta_{1}-x_{1}T      & a_{2} & b_{2} \\
x_{2} & y_{1}\Delta_{1}-x_{2}T & a_{3} & b_{3} \\
x_{3} & y_{2}\Delta_{1}-x_{3}T & a_{4} & b_{4}
\end{array}
\right)
\left(
\begin{array}{cccc}
\frac{1}{\sqrt{\Delta_{1}}} &  &                                  \\
      & \frac{1}{\sqrt{\Delta_{1}\Delta_{2}}} &                   \\
      &         & \frac{1}{\Delta_{1}\sqrt{\Delta_{2}\Delta_{3}}} \\
      &         &    & \frac{1}{\sqrt{\Delta_{3}}}
\end{array}
\right).
\end{eqnarray}

Similarly, starting from
\begin{equation}
F^{\prime}\equiv
\left(
\begin{array}{cccc}
1     & 0     & 0     & 0  \\
u_{1} & 1     & 0     & 0  \\
u_{2} & v_{1} & 1     & 0  \\
u_{3} & v_{2} & w_{1} & 1
\end{array}
\right)
\ \in\ GL(4;\fukuso)/B_{+}
\end{equation}
we have the unitary matrix parametrized by $(u_{1},u_{2},u_{3},
v_{1},v_{2},w_{1})$
\begin{equation}
\label{eq:}
U^{\prime}=
\left(
\begin{array}{cccc}
1     & -T^{\prime}                     & a_{1}^{\prime} & b_{1}^{\prime}          \\
u_{1} & \Delta_{1}^{\prime}-u_{1}T^{\prime}      & a_{2}^{\prime} & b_{2}^{\prime} \\
u_{2} & v_{1}\Delta_{1}^{\prime}-u_{2}T^{\prime} & a_{3}^{\prime} & b_{3}^{\prime} \\
u_{3} & v_{2}\Delta_{1}^{\prime}-u_{3}T^{\prime} & a_{4}^{\prime} & b_{4}^{\prime}
\end{array}
\right)
\left(
\begin{array}{cccc}
\frac{1}{\sqrt{\Delta_{1}^{\prime}}} &  &                                         \\
  & \frac{1}{\sqrt{\Delta_{1}^{\prime}\Delta_{2}^{\prime}}} &                     \\
  &  & \frac{1}{\Delta_{1}^{\prime}\sqrt{\Delta_{2}^{\prime}\Delta_{3}^{\prime}}} \\
  &  &  & \frac{1}{\sqrt{\Delta_{3}^{\prime}}}
\end{array}
\right)
\end{equation}
where $T^{\prime}=\bar{u}_{1}+\bar{u}_{2}v_{1}+\bar{u}_{3}v_{2}$ and etc.

As a result the CKM matrix $V=U^{\dagger}U^{\prime}$ is given by
\begin{eqnarray}
V&=&
\left(
\begin{array}{cccc}
\frac{1}{\sqrt{\Delta_{1}}} &  &                                  \\
      & \frac{1}{\sqrt{\Delta_{1}\Delta_{2}}} &                   \\
      &         & \frac{1}{\Delta_{1}\sqrt{\Delta_{2}\Delta_{3}}} \\
      &         &    & \frac{1}{\sqrt{\Delta_{3}}}
\end{array}
\right)
\left(
  \begin{array}{cccc}
    f_{11} & f_{12} & f_{13} & f_{14} \\
    f_{21} & f_{22} & f_{23} & f_{24} \\
    f_{31} & f_{32} & f_{33} & f_{34} \\
    f_{41} & f_{42} & f_{43} & f_{44}
  \end{array}
\right)\times \nonumber \\
&{}&
\left(
\begin{array}{cccc}
\frac{1}{\sqrt{\Delta_{1}^{\prime}}} &  &                                         \\
  & \frac{1}{\sqrt{\Delta_{1}^{\prime}\Delta_{2}^{\prime}}} &                     \\
  &  & \frac{1}{\Delta_{1}^{\prime}\sqrt{\Delta_{2}^{\prime}\Delta_{3}^{\prime}}} \\
  &  &  & \frac{1}{\sqrt{\Delta_{3}^{\prime}}}
\end{array}
\right)
\end{eqnarray}
where
\[
\left(
  \begin{array}{cccc}
    f_{11} & f_{12} & f_{13} & f_{14} \\
    f_{21} & f_{22} & f_{23} & f_{24} \\
    f_{31} & f_{32} & f_{33} & f_{34} \\
    f_{41} & f_{42} & f_{43} & f_{44}
  \end{array}
\right)
=
\left(
\begin{array}{cccc}
1     & -T                     & a_{1} & b_{1} \\
x_{1} & \Delta_{1}-x_{1}T      & a_{2} & b_{2} \\
x_{2} & y_{1}\Delta_{1}-x_{2}T & a_{3} & b_{3} \\
x_{3} & y_{2}\Delta_{1}-x_{3}T & a_{4} & b_{4}
\end{array}
\right)^{\dagger}
\left(
\begin{array}{cccc}
1     & -T^{\prime}                     & a_{1}^{\prime} & b_{1}^{\prime}          \\
u_{1} & \Delta_{1}^{\prime}-u_{1}T^{\prime}      & a_{2}^{\prime} & b_{2}^{\prime} \\
u_{2} & v_{1}\Delta_{1}^{\prime}-u_{2}T^{\prime} & a_{3}^{\prime} & b_{3}^{\prime} \\
u_{3} & v_{2}\Delta_{1}^{\prime}-u_{3}T^{\prime} & a_{4}^{\prime} & b_{4}^{\prime}
\end{array}
\right).
\]

\par \vspace{3mm} \noindent
This is our geometric parametrization to the CKM matrix in the fourth 
generation of qwarks. Though the form is a bit complicated, it is not 
avoidable.

\vspace{3mm}
A comment is in order.\ Jarlskog in \cite{Ja2}, \cite{Ja3} 
has given another parametrization to $SU(n)$, which is based on 
the canonical coordinate of the second kind in the Lie group theory. 
See also \cite{Fu2} and \cite{FFK}. 
However, the situation doesn't become simpler.

\section{Discussion}
In the paper we revisited the Kabayashi--Maskawa theory in the 
standard model from the geometric point of view and generalized 
some basic facts based on the third generation of quarks on 
CP--violation to ones based on the fourth generation. 

Though our method is of course not complete, to give a geometric 
insight to the standard model is very important.

To construct a unified model consisting of quarks and leptons we 
may treat a flag manifold $SU(6)/U(1)^{5}\cong U(6)/U(1)^{6}$, see 
for example \cite{GGPT}. In our method, starting from
\[
F\equiv 
\left(
\begin{array}{cccccc}
1     & 0     & 0     & 0     & 0     & 0 \\
u_{1} & 1     & 0     & 0     & 0     & 0 \\
u_{2} & v_{1} & 1     & 0     & 0     & 0 \\
u_{3} & v_{2} & w_{1} & 1     & 0     & 0 \\
u_{4} & v_{3} & w_{2} & x_{1} & 1     & 0 \\
u_{5} & v_{4} & w_{3} & x_{2} & y_{1} & 1
\end{array}
\right)
\ \in\ GL(6;\fukuso)/B_{+}
\]
we must perform the Gramm--Schmidt orthogonalization to obtain 
the unitary matrix $U$ likely in the text. However, the calculation  
becomes more and more hard. We will report it in the near future.

\vspace{5mm}
\noindent{\em Acknowledgment.}\\
K. Fujii wishes to thank Hiroshi Oike and Tatsuo Suzuki for 
their helpful comments and suggestion.

\vspace{10mm}
\begin{center}
 \begin{Large}
\noindent{\bfseries Appendix\quad Jarlskog Determinant}
 \end{Large}
\end{center}

\par \vspace{5mm} \noindent
In the appendix we calculate the determinant of the commutator 
$[M,M^{\prime}]$ in the general case. For mass matrices
\[
M=UDU^{\dagger},\quad M^{\prime}=U^{\prime}D^{\prime}{U^{\prime}}^{\dagger}
\]
where $D=\mbox{diag}(m_{1},m_{2},\cdots,m_{n})$ and 
$D^{\prime}=\mbox{diag}(m_{1}^{\prime},m_{2}^{\prime},\cdots,
m_{n}^{\prime})$, we want to calculate the Jarlskog determinant
\[
\det[M,M^{\prime}]
=\det(DVD^{\prime}V^{\dagger}-VD^{\prime}V^{\dagger}D)
\]
where $U^{\dagger}U^{\prime}\equiv V=(V_{ij})$ is the general CKM matrix. 

We set $X=DVD^{\prime}V^{\dagger}-VD^{\prime}V^{\dagger}D$ for simplicity. 
$X$ is anti--hermite ($X^{\dagger}=-X$), so
\[
\overline{\det(X)}=\det(X^{\dagger})=\det(-X)=(-1)^{n}\det(X).
\]
Therefore $\det(X)$ is {\bf real} if $n$ is even, while $\det(X)$ is 
{\bf pure imaginary} if $n$ is odd.

\par \noindent
Explicitly, 

\par \noindent
$n=2$ (real)
\[
\det(X)=(m_{2}-m_{1})^{2}(m_{2}^{\prime}-m_{1}^{\prime})^{2}
|V_{11}|^{2}|V_{21}|^{2},
\]

\par \noindent
$n=3$ (pure imaginary)
\begin{eqnarray*}
\det(X)&=&
(m_{3}-m_{1})(m_{3}-m_{2})(m_{2}-m_{1})
(m_{3}^{\prime}-m_{1}^{\prime})(m_{3}^{\prime}-m_{2}^{\prime})
(m_{2}^{\prime}-m_{1}^{\prime})\times \\
&{}&2i\ \mbox{Im}(V_{11}V_{22}\bar{V}_{12}\bar{V}_{21}).
\end{eqnarray*}

\par \noindent
$n=4$ (real)\quad The calculation is not easy. In \cite{Ja3} 
Jarlskog tries to calculate the term
\[
\mbox{Im}(V_{\alpha j}V_{\beta k}V_{\gamma l}
\bar{V}_{\alpha k}\bar{V}_{\beta l}\bar{V}_{\gamma j}),
\]
which is a ``natural" generalization when taking the case of $n=3$ 
into consideration. 
However, only such a term cannot be derived by the calculation 
above, \cite{Su}. We need further work.


\end{document}